\documentclass[%
 reprint,
 superscriptaddress,
 amsmath,amssymb,
 aps,
 prx,
]{revtex4-2}

\usepackage{graphicx}  
\usepackage{dcolumn}   
\usepackage{bm}        
\usepackage{hyperref}  
\usepackage{float}     
\usepackage{xcolor}    
\usepackage{siunitx}   
\usepackage{gensymb} 
\usepackage{ragged2e}  
\usepackage{breqn}

\makeatletter
\long\def\@makecaption#1#2{%
  \vskip\abovecaptionskip
  {\small \rmfamily \noindent {\bfseries #1.} \justifying #2\par}%
  \vskip\belowcaptionskip
}
\def\@float@makecaption{\@makecaption}
\makeatother

\hypersetup{
    colorlinks=true,
    linkcolor=blue,
    citecolor=blue,
    urlcolor=blue,
}

\begin{document}

\title{Fluid dynamics of a liquid mirror space telescope}

\author{Israel Gabay}
\affiliation{Faculty of Mechanical Engineering\begin{math},\end{math} Technion -- Israel Institute of Technology\begin{math},\end{math} Haifa\begin{math},\end{math} Israel.}

\author{Omer Luria}
\affiliation{Faculty of Mechanical Engineering\begin{math},\end{math} Technion -- Israel Institute of Technology\begin{math},\end{math} Haifa\begin{math},\end{math} Israel.}

\author{Edward Balaban}
\affiliation{NASA Ames Research Center\begin{math},\end{math} Moffett Blvd.\begin{math},\end{math} Moffett Field\begin{math},\end{math} CA\begin{math},\end{math} USA.}

\author{Amir D. Gat}
\email{amirgat@technion.ac.il}
\affiliation{Faculty of Mechanical Engineering\begin{math},\end{math} Technion -- Israel Institute of Technology\begin{math},\end{math} Haifa\begin{math},\end{math} Israel.}

\author{Moran Bercovici}
\email{mberco@technion.ac.il}
\affiliation{Faculty of Mechanical Engineering\begin{math},\end{math} Technion -- Israel Institute of Technology\begin{math},\end{math} Haifa\begin{math},\end{math} Israel.}

\date{\today}

\begin{abstract}
Large aperture telescopes are pivotal for exploring the universe, yet even with state-of-the-art manufacturing and launch technology, their size is limited to several meters. As we aim to build larger telescopes - extending tens of meters - designs in which the main mirror is based on liquid deployment in space are emerging as promising candidates. However, alongside their enormous potential advantages, liquid-based surfaces present new challenges in material science, mechanics, and fluid dynamics. One of the fundamental questions is whether it is possible for such surfaces to maintain their precise optical shape over long durations, and in particular under the forces induced by the telescope's accelerations. In this paper, we present a model and a closed-form analytical solution for the non-self-adjoint problem of the dynamics of a thin liquid film pinned within a finite circular domain. We use the 50-meter Fluidic Telescope (FLUTE) concept as the case study, and examine the liquid dynamics of the telescope under both slewing actuation and relaxation regimes, elucidating the role of geometrical parameters and liquid properties. The solutions reveal a maneuvering ‘budget’ wherein the degradation of the mirror surface is directly linked to the choice of maneuvers and their sequence. By simulating ten years of typical operation, we show that, while the maximal deformation might reach several microns, the spatial distribution of the deformation and their propagation rate allows the telescope to maintain its optical functionality for years, with at least a substantial portion of the aperture remaining suitable for astronomical observations. The model provides valuable insights and guidelines into the performance of liquid-film space telescopes, marking a crucial step toward realizing the potential of this innovative concept.
\end{abstract}

\maketitle

\section{Introduction}

Are we alone in the universe? This question captivates astrophysicists and space researchers worldwide~\cite{Ref1,Ref2}. Space telescopes with apertures of tens of meters in diameter (and thus capable of sufficiently detailed observations of extrasolar planets) could be a key tool in solving this mystery~\cite{Ref3,Ref4,Ref5}. However, despite significant advances in space telescope technologies, constructing such large  apertures remains extremely challenging~\cite{Ref6,Ref7}.

The largest operational space telescope, the James Webb Space Telescope (JWST), launched in 2021, has a primary mirror of 6.5 meters in diameter and yields an order of magnitude improvement in light collection compared to the previous flagship telescope, the Hubble Space Telescope~\cite{Ref8,Ref9}. JWST is one of the most expensive single-piece engineering projects in the world. For this mission, NASA used a launch vehicle with one of the largest payload bays available (Ariane 5)~\cite{Ref10,ariane5_jwst_prep}, in combination with the segmented main mirror, which was launched folded and then deployed in space. However, the segmentation approach cannot be realistically scaled to tens of meters~\cite{Ref6,Ref11}.

In recent years, there has been growing interest in liquid-based optical components~\cite{Ref12,Ref13,Ref14} for both ground and space telescopes, due to the nature of liquid interfaces which provides smooth and continuous surfaces. The concept of liquid-based ground telescope dates back to Newton~\cite{Ref15,Ref16} who suggested that a rotating liquid metal tank with a base perpendicular to the gravitational field would shape the liquid interface into a paraboloid surface~\cite{Ref17,Ref18}. Technological challenges prevented Newton from implementing such a telescope, and it became a reality only in the late 19th century~\cite{Ref16,Ref19}. Although the optical quality of such telescopes can be good, they suffer from several major limitations. First and foremost, such a telescope can point only in the zenith direction (normal to the surface of the earth) and cannot be slewed without losing its parabolic shape and thus its optical functionality. A second limitation is on size, as accurately controlling the rotation of larger mirrors becomes a significant technological challenge. While operation by rotation requires gravity and thus is more suitable for ground telescopes (either on Earth, or e.g., on the moon), it is also conceivable to implement rotating space telescope where the role of gravity is achieved by acceleration along the rotation axis.

The Zenith project~\cite{Ref12,Ref20,Ref21}, led by the Defense Advanced Research Projects Agency (DARPA), aims to create slewable Earth-based liquid mirror telescopes. The general approach taken by all the participating teams so far is control of the mirror surface through electromagnetic fields. Recently Rowlands et al.~\cite{Ref21} presented their steady-state model for the topography of a liquid telescope under magnetic actuation which induces normal stresses at the liquid-fluid interface. By assuming that the liquid film reachs a static hydro-magnetic state they solve for the liquid topography under the influence of gravity, surface tension, and the Maxwell stress resulting from the magnetic fields.

The FLUTE (Fluidic Telescope) project~\cite{Ref22,Ref23,Ref24}, which the authors of this paper are associated with, takes a different approach to creating a liquid-mirror telescope. The project---a joint effort between NASA and Technion---has a goal of creating space telescopes with mirrors measuring in tens of meters in diameter. In the FLUTE approach, the reflecting surface is created with a thin liquid film which, in microgravity, will naturally acquire a spherical shape corresponding to its minimum energy state. Frumkin et al.~\cite{Ref25} and Elgarisi et al.~\cite{Ref26} have provided analytical solutions for the steady-state shape of such a film, as a function of its liquid volume and boundary conditions. Based on this theoretical framework, one can consider surface corrections for the minimum energy surface through active liquid actuation. However, even in the absence of such actuation, the FLUTE approach must account for liquid dynamics as the telescope slews to different astronomical targets and performs orbit maintenance maneuvers, thus subjecting the liquid to perturbations.

Regardless of the specific approach, from a fluid mechanics perspective, the problem of a liquid-based space telescope can be considered as a finite volume of liquid maintained within a circular domain. While steady-state computations are an important first step in understanding the potential of such systems, film dynamics cannot be neglected when considering the actual operation of actuation devices. We provide, for the first time, a theoretical model and an analytical solution for film dynamics in a finite circular domain. We use the FLUTE approach as the case study of space-based liquid mirror dynamics and study the telescope’s response to one of the most significant sources of perturbations for a space telescope--its own slewing maneuvers required for changing its line of sight in space. We use this model to gain insight into the liquid dynamics resulting from such maneuvers and offer simplified expressions to predict film deformations based on geometrical parameters, liquid properties, and maneuver characteristics, using a non-dimensional analysis of the actuation and relaxation stages. Furthermore, we simulate ten years of operation of such a hypothetical telescope and show that, while the maximum deformation can reach thousands of nanometers (for a telescope \SI{50}{\meter} in diameter with a \SI{1}{\milli\meter} film thickness), the rate of propagation of these disturbances toward the center of the telescope remains slow enough to allow for years of operation with a substantial portion of its aperture remaining suitable for high-quality astronomical observations.

\section{The fluidic telescope model}

\begin{figure}[htbp]
\centering
\includegraphics[width=0.8\columnwidth]{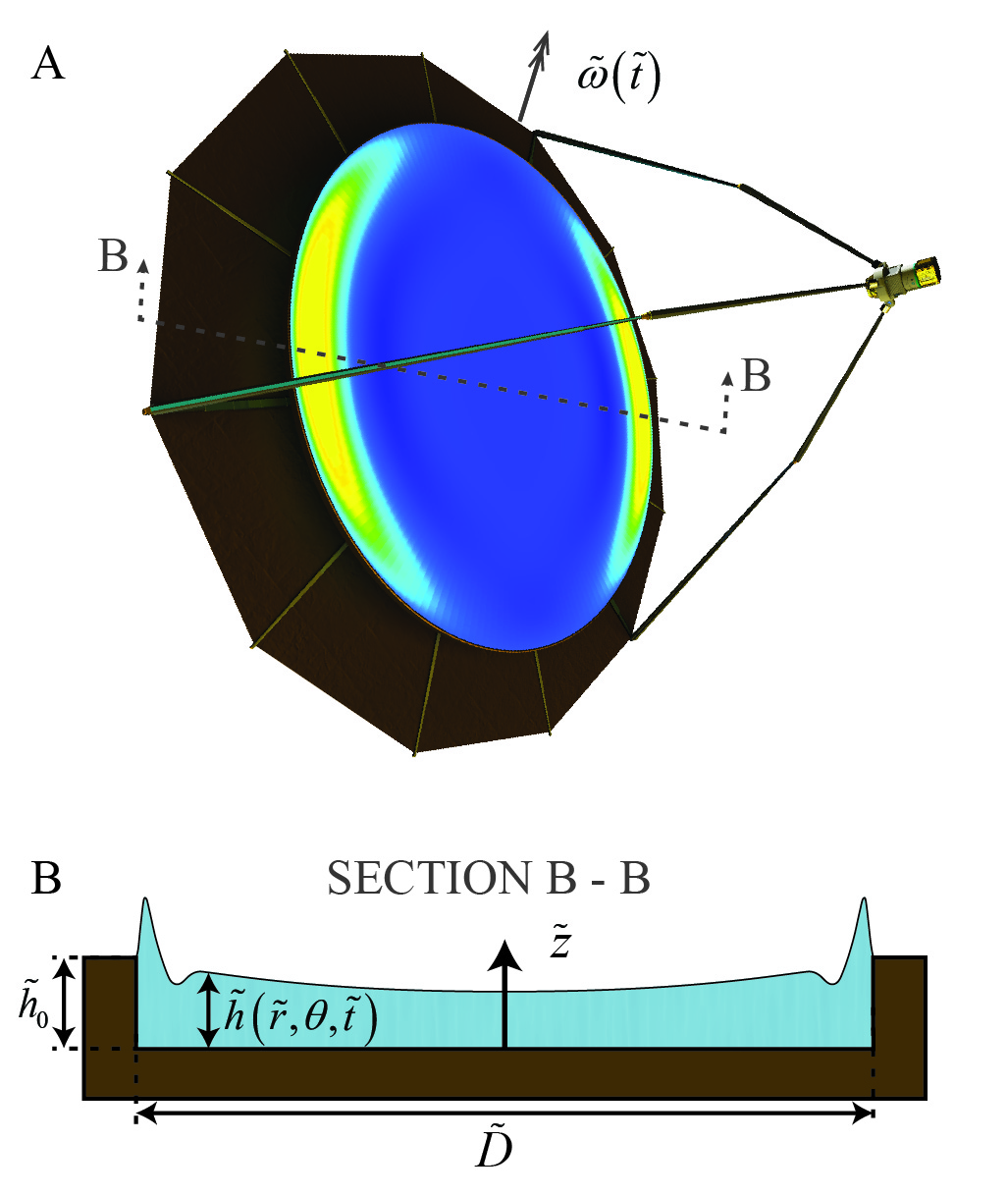}
\caption{\label{fig:Fig1}Schematic illustration of the investigated problem. (A) Artist impression of the FLUTE telescope in space undergoing a slewing maneuver at an angular velocity $\tilde{\omega}$. The telescope's primary mirror is made of a thin liquid film that takes a nominally spherical shape but deforms due to the maneuver, as indicated by the colormap. (B) A cross-section of the considered model. We model the telescope as a thin liquid film contained within a circular frame and pinned at its inner edge, and seek to solve for the deformation $\tilde{h}$ as a function of the polar coordinates $\tilde{r}$, $\theta$ and time $\tilde{t}$. The thickness of the liquid and its deformation are drawn not to scale, the film thickness is typically 4 orders of magnitude smaller than its diameter, and the deformations are four-six orders of magnitude smaller than its thickness.}

\end{figure}
Figure~\ref{fig:Fig1} illustrates the concept of the liquid-based space telescope. The telescope is modeled as a circular chamber with diameter $\tilde{D}$ and height $\tilde{h}_0$ filled with liquid of mass density $\tilde{\rho}$ and dynamic viscosity $\tilde{\mu}$. We consider slewing maneuvers, meaning that the telescope is subjected to an angular velocity $\tilde{\omega}$ around an in-plane rotation axis that passes through the center of the dish as shown in Figure~\ref{fig:Fig1}A. Under the assumption of shallow geometry, $\tilde{h}_0 \ll \tilde{D}$, and negligible fluid inertia $\text{Re}_r \ll 1$, the fluid momentum equations reduce to the lubrication equations,

\begin{subequations}
\begin{align}
\tilde{\nabla}_s \tilde{p} &= \tilde{\mu} \frac{\partial^2 \tilde{\mathbf{u}}_s}{\partial \tilde{z}^2} - \tilde{\rho} \tilde{\mathbf{a}}_s, \label{eq:1a} \\
\frac{\partial \tilde{p}}{\partial \tilde{z}} &= -\tilde{\rho} \tilde{g} \tilde{z} - \tilde{\rho} \tilde{a}_z, \label{eq:1b}
\end{align}
\end{subequations}

while the continuity equation (mass conservation) can be written as
\begin{equation}
\tilde{\nabla}_s \cdot \tilde{\mathbf{u}}_s + \frac{\partial \tilde{u}_z}{\partial \tilde{z}} = 0. \label{eq:2}
\end{equation}

The subscripts $s$ and $z$ mark the parallel and the transverse directions to the chamber's floor respectively, and $\tilde{\mathbf{a}}$ is the coordinate system acceleration. For simplicity of the model and its solution, and without loss of generality, we consider only rotation of the telescope around the $\tilde{y}$ axis, $\tilde{\boldsymbol{\omega}} = \tilde{\omega}(\tilde{t}) \hat{\mathbf{y}}$. Thus, the system acceleration can be written as
\begin{equation}
\tilde{\mathbf{a}} = -2\tilde{\omega}\tilde{u}\hat{\mathbf{x}} + 2\tilde{\omega}\tilde{v}\hat{\mathbf{x}} - \tilde{\omega}^2 \tilde{x}\hat{\mathbf{x}} - \tilde{\omega}^2 \tilde{z}\hat{\mathbf{z}} - \tilde{\omega}\tilde{x}\hat{\mathbf{z}} + \tilde{\omega}\tilde{z}\hat{\mathbf{x}}. \label{eq:3}
\end{equation}

Substituting the system acceleration into the momentum equations yields
\begin{subequations}
\begin{align}
\tilde{\nabla}_s \tilde{p} &= \tilde{\mu} \frac{\partial^2 \tilde{\mathbf{u}}_s}{\partial \tilde{z}^2} + \tilde{\rho} \left( -2\tilde{\omega}\tilde{v} - \tilde{\omega}\tilde{z} + \tilde{\omega}^2 \tilde{x} \right) \hat{\mathbf{x}}, \label{eq:4a} \\
\frac{\partial \tilde{p}}{\partial \tilde{z}} &= \tilde{\rho} \left( \tilde{\omega}^2 \tilde{z} + 2\tilde{\omega}\tilde{u} + \tilde{\omega}\tilde{x} - \tilde{g} \right) \hat{\mathbf{z}}. \label{eq:4b}
\end{align}
\end{subequations}

Under the negligible fluid inertia limit (i.e., small reduced-Reynolds number), the telescope slewing velocity must satisfy, $\tilde{\omega}_0 \ll \tilde{\nu} / \tilde{h}_0^2$. This upper bound ensures that the Coriolis terms in the momentum equations are negligible as well. This can be seen by comparing the magnitude of the Coriolis terms to that of the centrifugal terms in the in-plane momentum equation, (\ref{eq:4a}). Similarly, placing an upper bound on the angular acceleration, $\dot{\tilde{\omega}} \ll \tilde{\omega}_0^2 / \varepsilon$, ensuring that the Euler terms in both momentum equations are negligible as well. Thus, by considering a working regime where both the angular velocity and acceleration remain within the bounds established above, we obtain the simplified momentum equations,

\begin{subequations}
\begin{align}
\tilde{\nabla}_s \tilde{p} &= \tilde{\mu} \frac{\partial^2 \tilde{\mathbf{u}}_s}{\partial \tilde{z}^2} + \tilde{\rho} \tilde{\omega}^2 \tilde{x}\hat{\mathbf{x}}, \label{eq:5a} \\
\frac{\partial \tilde{p}}{\partial \tilde{z}} &= -\tilde{\rho} \tilde{g} \tilde{z}, \label{eq:5b}
\end{align}
\end{subequations}

where the dominant forcing term from the slewing is the centrifugal force. To derive the evolution equation representing the film height dynamics, we integrate the in-plane momentum equation, (\ref{eq:5a}), with respect to $\tilde{z}$ twice, and apply the no-slip and no tangential stress conditions on the lower and upper boundaries respectively,
\begin{equation}
\tilde{\mathbf{u}}_s = \frac{1}{2\tilde{\mu}} \left( \tilde{\nabla}_s \tilde{p} - \tilde{\rho} \tilde{\omega}^2 \tilde{x}\hat{\mathbf{x}} \right) \left( \tilde{z}^2 - 2\tilde{z}\tilde{h} \right). \label{eq:6}
\end{equation}

Solving the perpendicular momentum equation, (\ref{eq:5b}), and using the normal stress balance at the free surface, we obtain the pressure distribution in the liquid film,
\begin{equation}
\tilde{p} - \tilde{p}_0 = -\tilde{\sigma} \tilde{\nabla}^2 \tilde{h} + \tilde{\rho} \tilde{g} \tilde{h} - \tilde{\rho} \tilde{g} \tilde{z}, \label{eq:7}
\end{equation}
where $\tilde{p}$ and $\tilde{p}_0$ are the pressure of the liquid and the environment, respectively. Using the kinematic condition, $\tilde{h}_{\tilde{t}} + \tilde{\nabla}_s \cdot \int_0^{\tilde{h}} \tilde{\mathbf{u}}_s d\tilde{z} = 0$, and substituting the velocity profile and liquid pressure we obtain the evolution equation,
\begin{equation}
\tilde{h}_{\tilde{t}} = \frac{1}{3\tilde{\mu}} \tilde{\nabla}_s \cdot \left[ \tilde{h}^3 \left( \tilde{\nabla}_s \left( -\tilde{\sigma}\tilde{\nabla}^2\tilde{h} + \tilde{\rho}\tilde{g}\tilde{h} \right) - \tilde{\rho}\tilde{\omega}^2 \tilde{x}\hat{\mathbf{x}} \right) \right]. \label{eq:8}
\end{equation}

By further assuming small deformations (compared to the liquid mirror thickness), $\tilde{d} \ll \tilde{h}_0$, and defining the following non-dimensional variables, $\tilde{t} = \tilde{\tau}t$, $\tilde{r} = r\tilde{D}/2$, $\tilde{d} = d\tilde{h}_0$, we obtain the linearized deformation equation,
\begin{equation}
d_t + \nabla_s^4 d - Bo\nabla_s^2 d = -A\omega^2 \left( t \right), \label{eq:9}
\end{equation}
where we define $\tilde{\tau} = \frac{3\tilde{\mu}\tilde{D}^4}{16\tilde{h}_0^3 \tilde{\sigma}}$, $Bo = \frac{\tilde{\rho} g \tilde{D}^2}{4\tilde{\sigma}}$, and $A = \frac{\tilde{\rho}\tilde{\omega}_0^2 \tilde{D}^4}{16\tilde{\sigma}\tilde{h}_0}$. The second term of the LHS of the non-dimensional evolution equation (the capillary forces term) is negligible compared to the source term, yet we must account for it, as it is the highest order term, and thus is needed to satisfy the boundary conditions. Consequently, near the edges, there is a narrow boundary-layer region in which the capillary forces are important.

As shown in Figure 1, we examine a closed chamber with rigid walls at its boundaries. Due to the chamber geometry, the film must be pinned at the edges, meaning $\tilde{h} = \tilde{h}_0$ at the boundary. Additionally, due to the rigid walls at the chamber perimeter, the in-plane velocity, $\tilde{\mathbf{u}}_s = 0$, must vanish at the boundary. Thus, integrating the velocity profile, equation (\ref{eq:6}), with respect to $\tilde{z}$, evaluating the expression at the boundary, and taking only the radial component gives:
\begin{equation}
\tilde{p}_{\tilde{r}} = \tilde{\rho}\tilde{\omega}^2 \tilde{r}^2 \cos^2 \theta. \label{eq:10}
\end{equation}

Substituting this pressure condition into equation (\ref{eq:7}), the liquid's pressure, yields the no penetration boundary condition in terms of the film height, thus the system boundary conditions, at $\tilde{r} = \tilde{D}/2, \theta, \text{ and } \tilde{t} \geq 0$ are:
\begin{equation} \label{eq:11}
\begin{aligned}
&\tilde{h} = \tilde{h}_0, \\
&\tilde{h}_{\tilde{r}rr} - \frac{1}{\tilde{r}^2}\tilde{h}_r + \frac{1}{\tilde{r}}\tilde{h}_{r\theta} - \frac{1}{\tilde{r}^3}\tilde{h}_{\theta\theta} + \\ & \quad \quad \quad \quad + \frac{1} {\tilde{r}^2}\tilde{h}_{\theta\theta r} + \tilde{\rho}\tilde{g}\tilde{h}_r = -\frac{\tilde{\rho}\tilde{\omega}_0^2 \tilde{r}}{\tilde{\sigma}} \omega^2 (\tilde{t}) \cos^2 \theta 
\end{aligned}
\end{equation}
which in terms of the surface deformation for non-dimensional variables form the boundary conditions,
\begin{equation} \label{eq:12}
\begin{aligned}
&d = 0, \\
&d_{rr} + \left( Bo - 1 \right) d_r + d_{r\theta} - 2d_{\theta\theta} + d_{\theta\theta r} = -A \left( \omega(t) \cos \theta \right)^2 
\end{aligned}
\end{equation}
at $r = 1, \theta, \text{ and } t \geq 0$.
\subsection{Analytical solution}

The evolution of the film deformation is governed by a linear fourth-order partial differential equation, (\ref{eq:9}), and is subject to boundary conditions imposed at the chamber edge, (\ref{eq:12}), and an arbitrary initial deformation profile. Gabay et al. showed that a 1D pinned thin film problem in a closed chamber forms a non-self-adjoint problem \cite{Ref27}. The 2D case we study here has similar mathematical characteristics. This adds complexity to the solution procedure, particularly in constructing a complete basis for projection of arbitrary inputs. To address this, we employ a generalized separation of variables method adapted for non-self-adjoint systems \cite{Ref28}.

The first step is to homogenize the boundary conditions. The no-penetration condition at the chamber edge introduces azimuthal dependence and mixed derivatives, which preclude direct application of separation of variables. To overcome this, we define a new auxiliary variable, $\delta(r,\theta,t)$, such that
\begin{equation}
d(r,\theta,t) = \delta(r,\theta,t) + f(r,\theta)g(t), \label{eq:13}
\end{equation}
where $d(r,\theta,t)$ is the film deformation,
\begin{equation}
f(r,\theta) = \sum_{n=0}^{9} C_n r^n \frac{\cos(2\theta) + 1}{2} \label{eq:14}
\end{equation}
is the spatial auxiliary function, and $g(t) = \omega^2(t)$ is the time dependent part. These functions are chosen such that the boundary conditions for $\delta$ are homogeneous, and such that substituting this variable into the governing equation yields a right-hand side that is free of singularities in both the radial and azimuthal directions. The set of coefficients used to enforcing these requirements is,

\begin{equation} \label{eq:15}
\begin{aligned}
C_1 &= \frac{A(735 + 2Bo)}{10(504 + Bo)}, \quad C_2 = 0,  \\ 
C_3 &= \frac{A(-1260 + Bo(365 + Bo))}{80(504 + Bo)}, \\
C_4 &= \frac{7A(47376 + Bo(507 + Bo))}{96(504 + Bo)}, \quad C_5 = 0,\\
C_6 &= \frac{A(66150 + Bo(567 + Bo))}{6(504 + Bo)}, \\
C_7 &= -\frac{3A(367920 + Bo(2977 + 5Bo))}{80(504 + Bo)}, \\
C_8 &= \frac{5A(75852 + Bo(609 + Bo))}{48(504 + Bo)}, \\
C_9 &= -\frac{11A(75600 + Bo(615 + Bo))}{480(504 + Bo)}.
\end{aligned}
\end{equation}

Substituting this decomposition into the PDE system, equations (\ref{eq:9} and \ref{eq:12}), transforms the system into a homogeneous PDE for $\delta(r,\theta,t)$. Where the PDE is:
\begin{align}
\delta_t +\nabla_s^4 \delta - Bo\nabla_s^2 \delta &= -g_t f(r,\theta) \nonumber \\
&\quad - g(A + \nabla_s^4 f - Bo\nabla_s^2 f), \label{eq:16}
\end{align}
the boundary conditions at $r = 1, \theta, \text{ and } t \geq 0$ are:
\begin{equation} \label{eq:17}
\begin{cases}
&\delta = 0, \\
&\delta_{rr} + (Bo - 1)\delta_r + \delta_{rr} - 2\delta_{g\theta} + \delta_{\theta\theta} = 0
\end{cases}
\end{equation}
and an  initial condition
\begin{equation}\label{eq:18}
\delta(r,\theta,0) = d(r,\theta,0) - F(r,\theta,0).
\end{equation}

We then assume a separable solution of the form $\delta(r,\theta,t) = R(r)\Theta(\theta)T(t)$. We guess the azimuthal component solution for a periodic domain, with eigenfunctions $\Theta_n = a_n \cos(n\theta) + b_n \sin(n\theta)$. These form a complete orthogonal basis in the azimuthal direction. Substituting these into the linear PDE and dividing through by $\delta$ yields two ordinary differential equations, one in $r$ and $\theta$, and one in $t$, each equated to a constant separation parameter,

\begin{equation}\label{eq:19}
\begin{split}
-\frac{T'}{T} = \lambda^4 &= \frac{r^4 R^{(4)} + 2r^3 R^{(3)} - r^2 R'' + rR'}{Rr^4} \\
&\quad + \frac{-n^2(4R - 2rR' + r^2 R'' - Bor^2 R)}{Rr^4} \\
&\quad + \frac{-Bo(r^4 R'' + r^3 R')+ n^4 R}{Rr^4}.
\end{split}
\end{equation}
The radial component results in a fourth-order differential equation with coefficients that depend on $r$, involving combinations of Bessel and modified Bessel operators and can be presented in a more compact form:
\begin{equation} \label{eq:20}
L_2[L_1[R]] = 0,
\end{equation}
where $L_1 = \partial_{rr} + \frac{\partial_r}{r} - \left(\frac{n^2}{r^2} + \lambda^2\right)$, and $L_2 = \partial_{rr} + \frac{\partial_r}{r} - \left(Bo + \frac{n^2}{r^2} - \lambda^2\right)$. The eigenfunctions $R$ satisfy boundary conditions corresponding to zero deformation and zero net flow at the domain edge, along with regularity at the center,
\begin{subequations}
\begin{align}
&R(1) = 0, \label{eq:21a} \\
&R^{(3)}(1) + R''(1) - \left(1 + Bo + n^2\right)R'(1) = 0, \label{eq:21b} \\
&|R(0)| < \infty.  \label{eq:21c} 
\end{align}
\end{subequations}

Solving equation \eqref{eq:19} and applying these conditions yields the transcendental equation for the radial eigenvalues,
\begin{equation} \label{eq:22}
\alpha\beta^2 J_{n-1}(\alpha) = \left[\alpha^2 + \beta^2\right]n + \beta\alpha^2 \frac{I_{n-1}(\beta)}{I_n(\beta)} J_n(\alpha),
\end{equation}
and the eigenfunctions of the system,
\begin{equation} \label{eq:23}
R_{nm}(r) = \begin{cases}
J_n(\alpha_{nm}r) - \frac{J_n(\alpha_{nm})}{I_n(\beta_{nm})} I_n(\beta_{nm}r) &n \geq 0, m > 0 \\
0 &n > 0, m = 0, \\
\frac{I_0(\sqrt{Bo}r)}{1 - I_0(\sqrt{Bo})} &n = 0, m = 0
\end{cases}
\end{equation}
where $\alpha_{nm} = \sqrt{-Bo + \sqrt{Bo^2 + 4\lambda_{nm}^4}}$ and $\beta_{nm} = \sqrt{Bo + \sqrt{Bo^2 + 4\lambda_{nm}^4}}$.

Among the radial eigenfunctions, the eigenfunction of $n = m = 0$ corresponds to the steady-state solution. In the limit of zero Bond number (i.e., in the absence of gravity), the $n = m = 0$ eigenfunction is no longer valid and the actual steady-state solution takes the form of a parabolic profile. This is consistent with the spherical shape expected for a free interface in microgravity and follows naturally from the linearization procedure used to derive the governing equation.

To project arbitrary initial conditions or source terms onto the system eigenfunction in the case of a non-self-adjoint problem, we must construct the adjoint problem. While the azimuthal operator and boundary conditions (periodicity) are self-adjoint and the azimuthal eigenfunctions form an orthogonal set, the full system is not. In particular, the radial eigenfunctions are not orthogonal under the standard inner product of a cylindrical coordinate system, $\langle U,V \rangle_r = \int_0^1 UV r dr$, where $U$ and $V$ are two distinct radial eigenfunctions.

Thus, following the procedure presented in \cite{Ref27}, we should define the complementary boundary conditions, which are the boundary conditions of the adjoint problem. To do so, we use the bi-orthogonality relation \cite{Ref28}, $\langle L[R], Y \rangle_r = \langle R, L^*[Y] \rangle_r$, where $Y$ is an adjoint eigenfunction, and $L^*$ is the adjoint operator, which in this case is the same as the original operator as Bessel and Modified Bessel operators are self-adjoint. Now, by integrating the bi-orthogonality relation by parts over the domain, we find the boundary conditions of the adjoint system:
\begin{subequations}
\begin{align}
&Y'(1) = 0, \label{eq:24a} \\
&Y'''(1) - n^2 Y'(1) = 0, \label{eq:24b} \\
&|Y(0)| < \infty, \label{eq:24c}
\end{align}
\end{subequations}
which the first two resulted from the bi-orthogonality relation and the third is the regularity condition at the chamber's center. Solving the same differential operator, \eqref{eq:19}, but with the adjoint boundary conditions yields the adjoint eigenfunctions
\begin{equation} \label{eq:25}
Y_{nm} = \begin{cases}
0 &n > 0, m = 0 \\
J_n(\alpha_{nm}) + \Pi(n,m) I_n(\beta_{nm}) &n \geq 0, m > 0 \\
\text{const} &n = 0, m = 0\\
\end{cases}
\end{equation}
where $\Pi(n,m)=\frac{\alpha_{nm}}{\beta_{nm}} \frac{J_{n-1}(\alpha_{nm}) - J_{n+1}(\alpha_{nm})}{I_{n-1}(\beta_{nm}) + I_{n+1} (\beta_{nm})}$.

To construct the film deformation solution, we substitute the separation of variables expression,
\begin{equation} \label{eq26}
\delta(r,\theta,t) = \sum_{n=0}^{\infty} \sum_{m=0}^{\infty} R_{nm}(r)\Theta_n(\theta)T_{nm}(t),
\end{equation}
into equation \eqref{eq:16} and obtain the first-order differential equation for the temporal eigenfunctions,
\begin{dmath}\label{eq:27}
\sum_{n=0}^{\infty} \sum_{m=0}^{\infty} R_{nm}\Theta_n T'_{nm} + \sum_{n=0}^{\infty} \sum_{m=0}^{\infty} \nabla_z^4(R_{nm}\Theta_n)T_{nm} - Bo \sum_{n=0}^{\infty} \sum_{m=0}^{\infty} \nabla_z^2(R_{nm}\Theta_n)T_{nm} 
= \sum_{n=0}^{\infty} \sum_{m=0}^{\infty} R_{nm}\Theta_n a_{nm}(t) +  \sum_{n=0}^{\infty} \sum_{m=0}^{\infty} R_{nm}\Theta_n b'_{nm}(t)
\end{dmath}
where the RHS of equation \eqref{eq:16} and its initial condition are expressed in terms of a double series of the system eigenfunctions:
\begin{equation} \label{eq:28}
\begin{aligned}
\sum_{n=0}^{\infty} \sum_{m=0}^{\infty} R_{nm}\Theta_n a_{nm} &= -g\left(A + \nabla_s^4 f - Bo \nabla_s^2 f\right), \\
\sum_{n=0}^{\infty} \sum_{m=0}^{\infty} R_{nm}\Theta_n b_{nm} &= -g_t f(r,\theta) , \\
\sum_{n=0}^{\infty} \sum_{m=0}^{\infty} R_{nm}\Theta_n I_{nm} &= \delta(r,\theta,0).
\end{aligned}
\end{equation}

The main issue one should be aware of while solving a non-self-adjoint problem using this method, is expressing each one of the terms in \eqref{eq:27} as a series of the system eigenfunctions. In contrast to the self-adjoint case, to express an arbitrary function, $\phi(r,\theta)$, in terms of the system spatial eigenfunctions, here we use the adjoint eigenfunctions:
\begin{equation} \label{eq:29}
\begin{aligned}
\phi(r,\theta) = \sum_{n=0}^{\infty} \sum_{m=0}^{\infty} c_{nm} R_{nm}(r)\Theta_n(\theta), \\
c_{nm} = \frac{\int_0^{2\pi} \int_0^1 Y_{nm}\Theta_n \phi(r,\theta) r dr d\theta}{\int_0^{2\pi} \int_0^1 Y_{nm} R_{nm}(r)\Theta_n^2(\theta) r dr d\theta}.
\end{aligned}
\end{equation}

Thus, the solution for each time-dependent function is obtained in the same manner,
\begin{equation} \label{eq:30}
T_{nm} = \int_{s=0}^t e^{\lambda_{nm}^4(s-t)}\left(a_{nm}(s) - b_{nm}(s)\right) ds + I_{nm} e^{-\lambda_{nm}^4 t},
\end{equation}
where,
\begin{equation} \label{eq:31}
\begin{aligned}
&a_{nm}(t) = -\frac{\int_0^{2\pi} \int_0^1 Y_{nm}\Theta_n g(t)(A + \nabla_z^4 f - Bo \nabla_z^2 f) r dr d\theta}{\int_0^{2\pi} \int_0^1 Y_{nm} R_{nm}(r)\Theta_n^2(\theta) r dr d\theta} ,\\
&b_{nm}(t) = -\frac{\int_0^{2\pi} \int_0^1 Y_{nm}\Theta_n g_t(t) f(r,\theta) r dr d\theta}{\int_0^{2\pi} \int_0^1 Y_{nm} R_{nm}(r)\Theta_n^2(\theta) r dr d\theta}, \\
&I_{nm} = -\frac{\int_0^{2\pi} \int_0^1 Y_{nm}\Theta_n \delta(r,\theta,0) r dr d\theta}{\int_0^{2\pi} \int_0^1 Y_{nm} R_{nm}(r)\Theta_n^2(\theta) r dr d\theta}.
\end{aligned}
\end{equation}

\section{Experimental validation}

\begin{figure*}[t]
    \includegraphics[width=\textwidth]{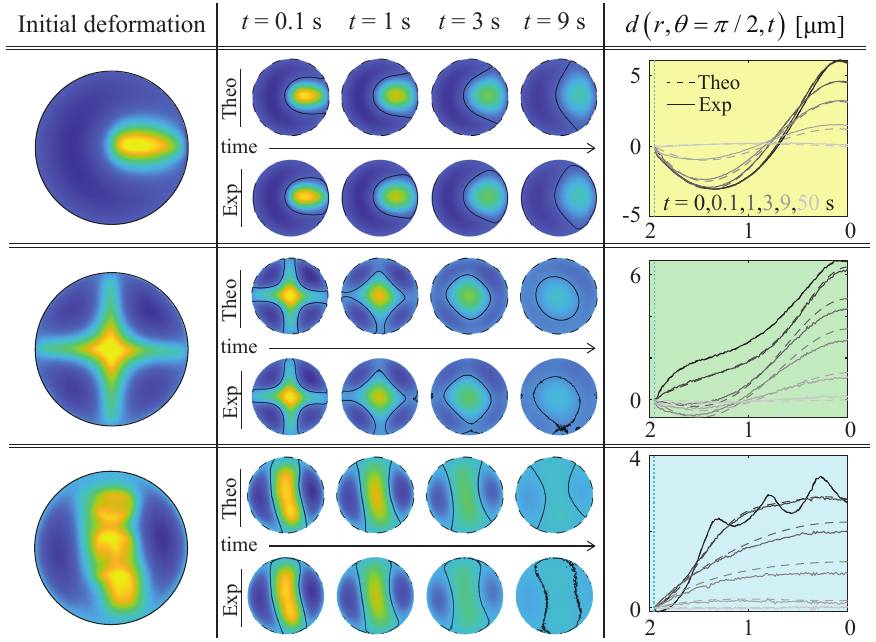} 
    \caption{\label{fig:Fig2} Experimental validation of the theoretical model for three different initial conditions. We deform the surface using DEP forces and use the resulting deformation as an initial condition to our model, which predicts the relaxation dynamics of the liquid-air interface. We compare the predictions to those measured in the experiment after the DEP forcing is released. The left column shows the initial deformation of each case. In the middle column, theoretical solution (upper row) and experimental measurements (lower row) are depicted for $t=0.1$~s, $1$~s, $3$~s, and $9$~s. The black contour in each map indicates the zero-deformation contour line. The right column presents a two-dimensional cross-section view along the radial direction at $\theta = \pi/2$, showing good agreement between the theoretical (dashed lines) and experimental (solid lines) deformation profiles at multiple times ($t=0$~s, $0.1$~s, $1$~s, $3$~s, $9$~s, $50$~s). Grayscale color coding differentiates the timing of measurements, with the initial deformation depicted in black, and later measurements in progressively lighter shades.}
\end{figure*}

Validating the mathematical model with a demonstration of a slewing telescope in conditions of microgravity was not practical. Therefore, we performed an alternative experiment under normal laboratory conditions (i.e., with gravity present), taking advantage of the fact that the mathematical model accounts for gravitational acceleration as a parameter. This allows direct comparison between the experimental results and theoretical predictions, even though the manuscript focuses on a space telescope intended for microgravity environments.

To validate the model in the lab setting, we sought a contact-free approach to impose an arbitrary initial deformation on the surface (one which is complex, and thus accounts for multiple modes), and observe its relaxation dynamics over time. To achieve this, we selected dielectrophoresis (DEP), which can induce a contactless force via electrodes located at the bottom of the film. The experimental setup consisted of a micro-fabricated circular chamber, with a diameter of \SI{3.8}{\milli\meter} and a height of \SI{50}{\micro\meter}, which was filled with \SI{0.57}{\micro\liter} silicone oil with a density of \SI{970}{\kilogram\per\meter\cubed} and a surface tension of \SI{20}{\milli\newton\per\meter}. DEP forces were applied through surface electrodes pre-patterned at the base of the chamber~\cite{Ref29}. 

Each experiment began by applying an electric field, allowing the liquid to stabilize in its new electro-hydrostatic state. When the voltage is turned off, the liquid undergoes relaxation, and we record the spatial deformation as a function of time using a digital holographic microscope (DHM Lyncee Tec R1000, 1.25X microscope objective).

Figure~\ref{fig:Fig2} illustrates the surface deformation from three distinct experiments, depicting three different initial shapes in the left column. In all cases, we allow the interface to reach a steady state, and all deformation data presented here is obtained by subtracting the steady state topography from the topography data of any other time point. In the middle column, interfacial deformation is plotted at four time points for each case, for both the experiments and the theoretical predictions. A contour line at $d=0$ is marked to guide the eye in comparison of the results. The right column offers quantitative comparisons by displaying cross-sections of the deformation maps at $\theta = \pi/2$, for both the theoretical (dashed line) and experimental (solid line) results. Throughout, the theoretical model shows good agreement with the experimental results.

\section{Liquid mirror deformation}

\begin{figure*}[t]
    \includegraphics[width=\textwidth]{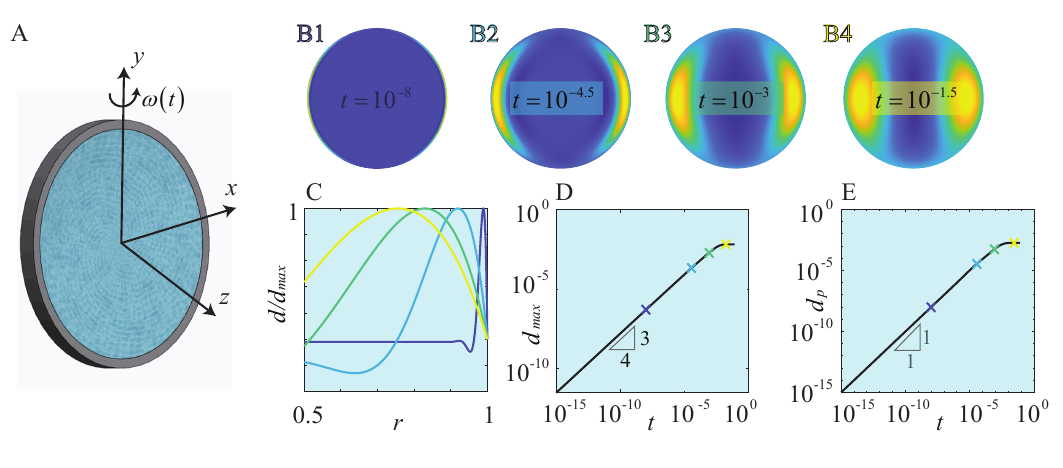}
    \caption{\label{fig:Fig3} Non-dimensional analysis of the liquid film’s deformation dynamics due to a continuous slewing maneuver. (A) Illustration of the telescope setup undergoing rotation around the y-axis. (B1–B4) The deformation map at four different times, which span 6.5 orders of magnitude, showing the progression of the deformation from the telescope edges (barely visible in B1) toward its center. (C) The deformation along the radial direction coinciding with the x-axis at the same time points depicted in B. (D–E) Maximal and piston deformations as a function of time on a logarithmic scale, identifying a scaling relation for the early time dynamics of \( d_{\mathrm{max}} \sim t^{3/4} \), and \( d_p \sim t^{-1/4} \) respectively. These relations hold for most of the actuation stage until late times. The ‘x’ marks represent time steps corresponding to the plotted deformation maps in B and C \( t = 10^{-8} \) dark blue, \( t = 4.5 \times 10^{-4} \) light blue, \( t = 10^{-3} \) green, and \( t = 1.5 \times 10^{-1} \) yellow.}
\end{figure*}

In the analytical model section, we provided a non-dimensional analytical solution of the film deformation. The non-dimensional representation is the most generic one, as it allows us to obtain insights onto the various time scales of the observed phenomenon, and its parametric representation allows computation for any desired configuration. However, it is also instructive to consider several specific cases, to demonstrate how the model could be used to predict the behavior of a hypothetical liquid telescope undergoing multiple maneuvers during its operation. Thus, in this section we provide both the non-dimensional investigation and several dimensional case studies.

\subsection{Non-dimensional investigation}

Figure~\ref{fig:Fig3} presents the deformation of the liquid mirror for the case of \( A = 1 \), resulting from a continuous slewing maneuver, \( \vec{\omega}(t) = \omega_0 \hat{y} \). As expected, the liquid is pushed towards the chamber edges along the x-axis under this type of actuation. As shown in panels B and C, at early times, the deformation builds up primarily within a narrow strip near the telescope's perimeter. The liquid is supplied from the center of the telescope, with the interface slowly receding while overall maintaining its spherical shape. We define this translation motion as ‘piston’ (commonly used in optics to describe a uniform phase bias) deformation, \( d_p \). Although such a movement of the surface will introduce optical aberrations, those can usually be corrected by translating components along the optical axis. For this reason, surface piston is treated here as a deformation which does not restrict the telescope’s performance. Panel D illustrates the maximum deformation as a function of time on a log-log scale, revealing a relation of \( d_{\mathrm{max}} \sim t^{3/4} \) throughout most of the actuation period, until a steady-state deformation, illustrated in panel B4, is achieved. Panel E shows the translation of the spherical surface maintained in an area within 80\% of the total radius.

\begin{figure}[t]
    \includegraphics{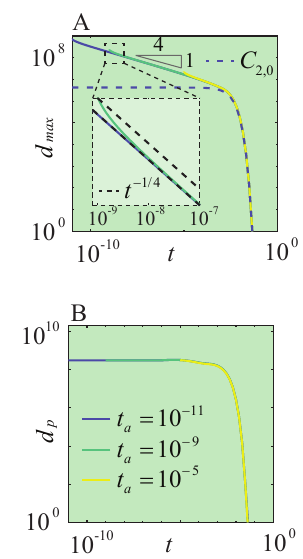}
    \caption{\label{fig:Fig4} The non-dimensional maximal and piston deformations of the liquid mirror as a function of time after the telescope stops slewing, i.e., the relaxation stage. Panels A and B illustrate the maximal and piston deformations, respectively, as a function of time (normalized by the deformation at the last time point, \( t = 0.03 \)) on a log-log scale. The different curves correspond to relaxations following actuations that began at \( t = 0 \) and ended at different times: \( t_a = 10^{-11} \) (blue), \( t_a = 10^{-9} \) (green), and \( t_a = 10^{-5} \) (yellow). (A) At the earliest times (inset) we observe rapid decay corresponding to the dynamics of the fastest modes (shortest wavelengths). At intermediate times all curves coincide, and the decay follows a \( t^{-1/4} \) dependence. At late times, the decay rate settles into that of the smallest wavenumber - the slowest mode \( (n = 2, m = 0) \). (B) The piston deformation is constant at both early and intermediate times and drops only when the higher modes decay and the shortest mode remains.}
\end{figure}

Figure~\ref{fig:Fig4} shows the maximal deformation and the piston deformation, rescaled by \( d_{\mathrm{max}}(t = 0.03) \) and \( d_p(t = 0.03) \), respectively, as a function of time during the relaxation stage, following a slewing actuation with a duration of \( t_a = 10^{-11} \) (blue line), \( t_a = 10^{-9} \) (green line), and \( t_a = 10^{-5} \) (yellow line). As illustrated in the figure, the relaxation stage of the liquid converges onto a single curve for both maximal and piston deformation, regardless of the duration of the actuation.

Panel A reveals three distinct regimes: immediately after the termination of the actuation, at very early times, we observe rapid decay corresponding to the highest wave numbers on the order of 20\% (as presented in the inset). At intermediate times (following the very early time and up to approximately \( 10^{-5} \)), the curves converge, and the decay of the maximum deformation follows a relation of \( d_{\mathrm{max}} \sim t^{-1/4} \), in agreement with the dynamics previously suggested by Zhang et al.~\cite{Ref30} for early-time behavior of healing capillary films. At later times, the decay of the maximum deformation corresponds to the dynamics of the longest wavelength (or the smallest eigenvalue) contributing to the deformation, which for the slewing maneuver type is the \( \cos(2\theta) \) term that dominates the source term in equation \eqref{eq:17}. This mode is depicted as the dashed line in panel A.

Under the FLUTE mission design concept, a telescope with a 50-meter primary mirror and a liquid film of several millimeters in thickness, with a reference viscosity of 100~cSt and surface tension of 50~mN/m, has a characteristic system timescale on the order of \( 10^{14} \). Thus, the entire operational time scale of a telescope, on the order of years, can be considered to be well within early- and intermediate-time regimes.

Using simulation results for both actuation and relaxation stages and the non-dimensional system parameters, we can provide predictions for the maximal deformation and piston deformation as functions of the telescope’s physical parameters. For the actuation regime, the maximal liquid deformation can be expressed as:
\begin{equation} \label{eq:32}
\tilde{d}_{\mathrm{max}}=d_{\mathrm{max}} \tilde{h}_0= A C_a t^{3/4} \tilde{h}_0 ,
\end{equation}
where \( C_a = 0.5192 \) is a constant calculated from Figure~\ref{fig:Fig4}D. By substituting the actuation time as a function of the angular velocity \( \tilde{\omega}_0 \), the requested slewing angle \( \beta \), and the system time scale \( \tilde{\tau} \), we obtain:
\begin{equation} \label{eq:33}
\tilde{d}_{\text{max},a} = \frac{C_a \tilde{\rho} \tilde{D} \tilde{\omega}_0^{5/4} \tilde{h}_0^{9/4}}{2\tilde{\gamma}^{1/4}} \left(\frac{\beta}{3\tilde{\mu}}\right)^{3/4}.
\end{equation}

Similarly, the piston deformation can be expressed as:
$\tilde{d}_p=d_p \tilde{h}_0 = -A t  \tilde{h}_0$ and by substituting the actuation time as a function of \( \tilde{\omega}_0 \), \( \beta \), and \( \tilde{\tau} \), we obtain:
\begin{equation}\label{eq:34}
\tilde{d}_{p,a} = -\frac{\beta \tilde{h}_0^3 \tilde{\rho} \tilde{\omega}_0}{3\tilde{\mu}}.
\end{equation}

For the relaxation stage, neglecting the short decay at early times, the decay of the maximal deformation can be estimated using the relation \( d_{\mathrm{max}} \sim t^{-1/4} \), i.e.,
\begin{equation} \label{eq:35}
\tilde{d}_{\mathrm{max}}(t) = \tilde{d}_{\mathrm{max},a} \left( \frac{t_a}{t} \right)^{1/4}, \quad t \geq t_a.
\end{equation}

where $t_a$ is the slewing time, and $\tilde{d}_{\mathrm{max},a}$ is the maximal deformation obtained at the end of the actuation stage.The piston deformation remains constant at early and intermediate times. This is simply because, at these time scales, the deformation that began at the edges did not yet penetrate into the center of the telescope.

\subsection{Dimensional investigation -- specific study cases}

Here we present different case studies considering variation in film height, angular velocity, and maneuvering procedures, e.g., several maneuvers, different maneuvering angles and variations in the axis of rotation. Yet, in all the cases considered here, we fix the properties of the liquid ($\tilde{\rho} = 1000$ kg/m$^3$, $\tilde{\nu} = 100$ cSt, $\tilde{\gamma} = 50$ mN/m), and the telescope diameter, $\tilde{D} = 50$ m.

Figure 5 presents the maximal deformation as a function of time during the relaxation phase for a full aperture (panel A) and a 90\% aperture (panel B) of the liquid mirror after performing a $45^\circ$ slewing maneuver in space. We investigate the influence of the film's thickness (ranging from 1 to 5 mm) and angular velocity (ranging from 10 to 100 $\mu$rad/s) on the deformation dynamics.

Since the slewing angle is identical across all configurations while the angular velocity varies, the actuation time differs between cases, causing the relaxation stage to begin at different times. The circles on each curve correspond to the deformation predicted by equation~\eqref{eq:33}, which match well with the simulation results of the full analytical solution~\eqref{eq:30}. The results also illustrate the $\tilde{\omega}_0^{5/4}$ and $\tilde{h}_0^{9/4}$ dependence of the maximal deformation predicted by equation~\eqref{eq:33}. The horizontal dashed line represents a strict imperfection limit in astronomy at a value of 20~nm, corresponding to $\tilde{\lambda}/20$, where $\tilde{\lambda} = 400$~nm is the lower edge of the visible wavelength---a common figure for high-quality space optics. All curves decay at approximately the same rate, consistent with the observation in Figure~\ref{fig:Fig4}. In cases 7--9, where the film thickness is larger, the system's time scale is shorter ($\tilde{\tau} \approx 2 \times 10^{13}$~s), making it possible to observe the transition into the sharper decay of the late-time regime at around $\tilde{t} = 100$~years. For the other cases, this transition occurs only at later times.

Interestingly, while all maneuvers exceed 20~nm at full aperture, when limiting the observation to a 90\% aperture (45-m diameter), the majority of cases remain below the 20-nm threshold. This occurs because the deformation originates at the telescope edges and propagates toward the center over time. The overall radial behavior of the deformation resembles an oscillating wave with diminishing amplitude toward the center. When observing with the full aperture, the maximal deformation decays monotonically, as it corresponds to the decay of the largest peak. However, the maximal deformation within the 90\% aperture corresponds to the peak value included within that region at any given time. Since the wave propagates inward, we observe patterns of alternating peak decay with sudden increases corresponding to the arrival of new peaks.

\begin{figure}[t]
\includegraphics{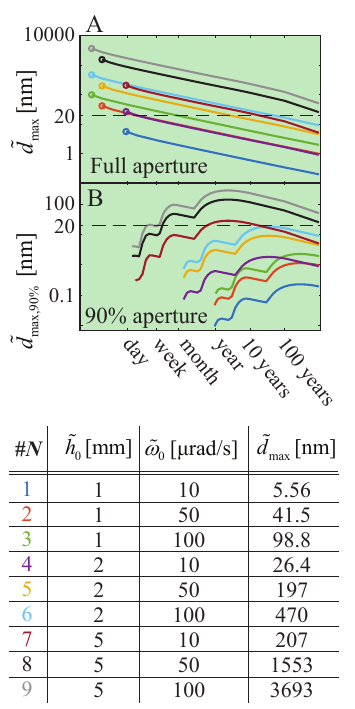}
\caption{Dimensional time-dependent deformation dynamics of a 50-m diameter liquid mirror during the relaxation phase, following a 45$^\circ$ slewing maneuver on a log-log scale. The table details the different experimental cases, specifying film thickness, angular velocity, and the resulting maximal deformation. Each row in the table corresponds to a curve in panels A and B, with matching colors to facilitate easy comparison. (A) The maximal deformation over time for a full aperture. The circles correspond to deformations predicted by equation~\eqref{eq:33}, aligning with the analytical solution. The horizontal dashed line at 20~nm serves as a reference, corresponding to $\tilde{\lambda}/20$ (where $\tilde{\lambda} = 400$~nm), an acceptable standard for high-quality space optics. The results show that regardless of the initial conditions, the deformation decays at the same rate at intermediate times. (B) The maximal deformation within a 90\% aperture (i.e., 45-m diameter). While for the full aperture, the majority of cases exceed the 20~nm threshold, for the 90\% aperture many cases remain below this mark for extended durations, since the deformations, while large, are confined to the outer regions of the telescope.}
\label{fig:Fig5}
\end{figure}

\begin{figure*}[t]
\includegraphics{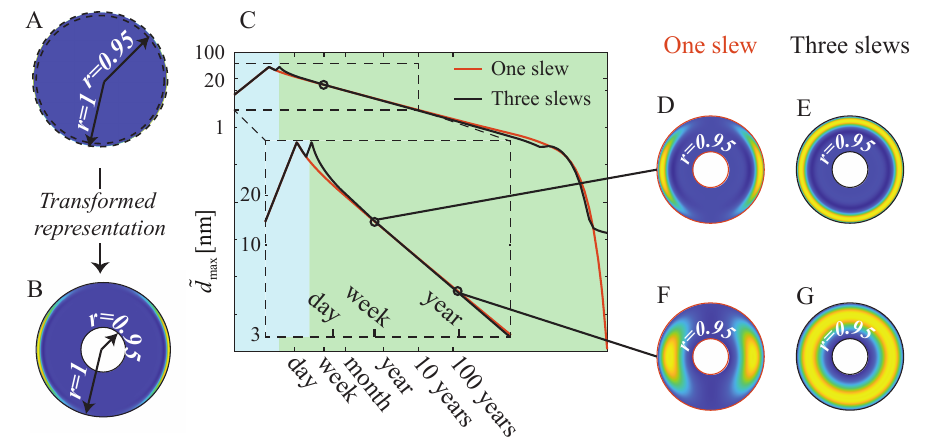}
\caption{Comparison of the liquid mirror deformation for two different sets of maneuvers of a 50~m diameter telescope with a 1~mm liquid layer. The first set involved a single 45$^\circ$ slewing maneuver around the $y$-axis. The second set incorporates the same 45$^\circ$ maneuver around the $y$-axis (as in the first set), followed by a subsequent 22.5$^\circ$ rotation around the $x$-axis and a $-22.5^\circ$ rotation back around the same axis. (A) The deformations of the liquid at a representative intermediate time, plotted over the full aperture. Since the deformation remains confined to a narrow region along the frame's edges it is not clearly visible in this representation. (B) Hereafter, we use a transformed graphical representation focusing only on the outer 5\% of the telescope radius, with the inner 95\% deliberately omitted. (C) The maximal deformation dynamics for the two cases, on a logarithmic time scale. The actuation period is represented with a blue background, and the relaxation phase with a green background. The inset focuses on the first few years to better illustrate the difference between the two sets of maneuvers investigated. Interestingly, since the additional maneuver occurs in a perpendicular direction, it does not increase the maximal deformation beyond its value in the first maneuver. (D--G) The surface deformation of the outer 5\% of the telescope radius after one week (D and E) and one year (F--G). While the maximal deformation in both is similar, the three-slews maneuver results in a more axisymmetric deformation.}
\label{fig:Fig6}
\end{figure*}

We now turn to examine the effects of multiple sequential maneuvers on the deformation dynamics. For clarity and simplicity, hereafter we also fix the film's thickness to be $\tilde{h}_0 = 1$~mm and the angular velocity to be  $\tilde{\omega}_0 = 50 \mu$rad/s (the conditions of configuration \#2 in Figure~5), and construct maneuver sequences from variations in the axis and magnitude of rotation. Within the time scales of interest, the deformation is limited to a narrow region close to the frame edges and cannot be discerned when plotting the deformation map of the entire telescope surface, as shown in Figure~6A. We thus adopt a transformed graphical representation of the surface that focuses on the outer 5\% of the telescope's radius, as depicted in Figure~6B. The area contained within the inner 95\% of the telescope's radius is purposely omitted and appears in white. In Figure~6C we compare the dynamics of the maximal deformation for two maneuver sequences: the first (in orange) is a single slewing maneuver of 45$^\circ$ around the $y$-axis. The second (in dashed black) is the same single slewing maneuver of 45$^\circ$ around the $y$-axis, followed immediately by a 22.5$^\circ$ maneuver around the $x$-axis and another $-22.5^\circ$ rotation back around the same axis---i.e., in both sequences the final position of the telescope is rotated by 45$^\circ$ around its original $y$-axis. The actuation period is shown on a blue background, whereas the relaxation phase has a green background.

The first sequence behaves as one would expect from the results of Figures~\ref{fig:Fig3} and \ref{fig:Fig4} - we see rapid build-up of the deformation to a level of 41 nm within approximately 4.5 hours, followed by a slow decay wherein the deformation reduces to 14 nm within a week and to 5 nm within a year. However, as shown in Figure~\ref{fig:Fig5} and depicted in the deformation map of Panels \ref{fig:Fig6}D,~\ref{fig:Fig6}F, the deformation remains confined to the outer ring and does not yet penetrate the central aperture of the telescope.

The results of the second sequence, which consists of maneuvers around multiple axes, show that simple superposition cannot be applied in such cases. When the first maneuver, shared by the two sequences, is complete (the orange line begins to decay), the second maneuver (around the perpendicular axis) is initiated. Surprisingly, despite the active actuation, the deformation does not continue growing and instead exhibits a decay that follows closely that of the first sequence for approximately 2~hours. This is because the rotation around this perpendicular axis drives deformations in regions that were previously unperturbed. Two processes now occur simultaneously---the decay of the initial deformation and the build-up of the new (perpendicular) deformation. Initially, we see a decay of the maximal deformation since the new deformation has not yet increased sufficiently to change the global maxima on the surface. However, with time, the growing new deformation will surpass the decaying first deformation and will dictate the global maximum, again resulting in an increase of the maximal deformation. After one week of decay, the maximal deformation of the two cases is nearly identical, yet, as shown in Figure~\ref{fig:Fig6}E, the surface of the second sequence is nearly axisymmetric, and this shape is maintained even at long times as shown in Figure~\ref{fig:Fig6}G. From this point onward, the two cases continue decaying at approximately the same pace, with a very slight advantage to the second sequence that decays marginally faster. Hence, from an optical perspective, these results suggest that it may sometimes be more beneficial to add maneuvers in order to shape the surface into a geometry that is easier to correct optically.

\begin{figure*}[t]
\includegraphics{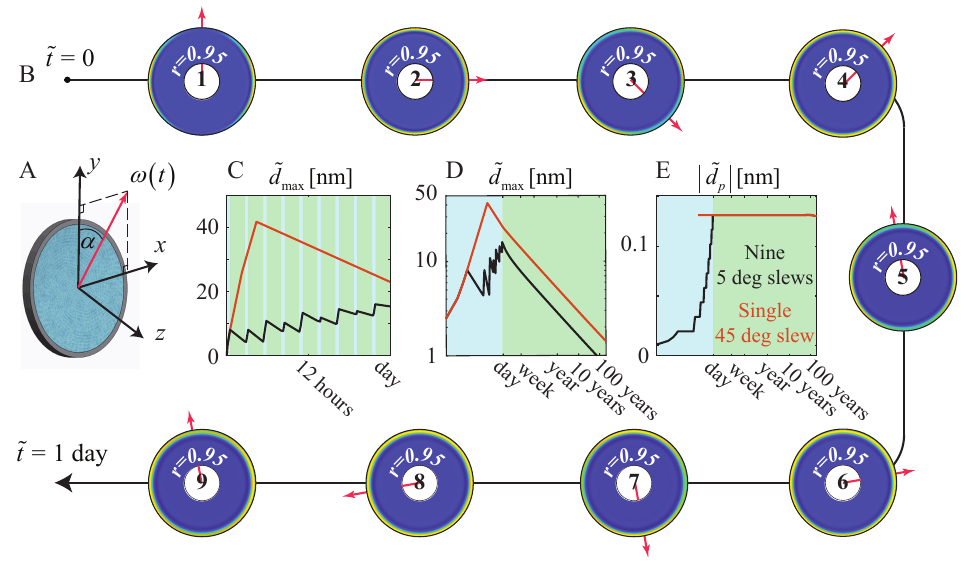}
\caption{Deformation dynamics of the 50~m diameter liquid mirror telescope over 24 hours as a result of 9 slewing maneuvers of 5$^{\circ}$ each around a randomly selected axis of rotation. (A) The axis of rotation (red arrow) is in the $x$-$y$ plane and is defined by the angle relative to the $y$-axis. (B) The deformation map after each of the 9 slewing maneuvers, showing the deformation propagating toward the center. The axis of rotation of each maneuver is presented as a red arrow for each step. (C) The black curve corresponds to the deformation due to the 9 maneuvers, showing the build-up and relaxation of deformation with each maneuver. As expected from Figure~\ref{fig:Fig6}, some rotations contribute minimally to the overall deformation. The orange curve presents, for comparison, the deformation due to a single 45$^{\circ}$ actuation and shows that multiple small maneuvers (in different directions) result in a smaller overall deformation. (D) The decay of the maximal deformation over longer time periods for the two cases. The decay rate is similar for both, with the initial difference in deformation persisting over time. (E) The piston deformation for both cases, noting that the incremental addition from each maneuver adds up to equal the total deformation seen in the single maneuver case.}
\label{fig:Fig7}
\end{figure*}

The JWST performs approximately 9 slewing maneuvers of several degrees each day~\cite{Ref31}. Using this as reference, Figure~\ref{fig:Fig7} shows the deformation of the liquid telescope within a 24-hour period, as it undergoes a set of 9 maneuvers, each of 5$^{\circ}$, around a random rotation axis defined by the angle $\alpha$ in its $x$-$y$ plane, as depicted in Figure~\ref{fig:Fig7}A. Each maneuver lasts approximately 30 minutes and is then followed by approximately 2 hours of observation time at which the telescope is at rest. Figure~\ref{fig:Fig7}C presents the deformation during the first 24 hours of actuation, showing the sequence of deformation build-up and relaxation between maneuvers. Depending on the direction of rotation, that is randomly assigned, we see that some actuations do not contribute significantly to the maximal deformation, as elucidated in Figure~\ref{fig:Fig6}. For comparison, we again present case \#2 of Figure~\ref{fig:Fig5} -- a single 45$^{\circ}$ maneuver over a 4.5-hour period. Interestingly, the maximal deformation of the two cases after one day is comparable. Figure~\ref{fig:Fig7}D presents the decay of the two cases over longer periods of time. After one day, the single maneuver case has already completed its rapid early decay phase and is entering the intermediate decay phase of $d_{\text{max}} \sim t^{-1/4}$. The multiple maneuvers case, on the other hand, is just entering its early decay phase, and thus decays more rapidly at that point. As a result, when both cases are well within their intermediate decay regime, the difference in maximal deformation between the two cases becomes more pronounced compared to the difference observed after one day.

Figure~\ref{fig:Fig7}E presents piston deformations for the same cases and time periods. The piston deformation builds up with each maneuver and remains constant thereafter. The additional deformation in each maneuver is independent of the axis of rotation, and the total piston deformation can be simply summed up for the nine-maneuver case and is equal to that obtained from the single maneuver case. Figure~\ref{fig:Fig7}B presents the deformation map of the mirror (within the 95\% outer radius region), after each of the slewing maneuvers, and the axis of rotation for each step (red arrow). We see the deformation propagating toward the center, and, with time, becoming less sensitive to variations in the axis of rotation. Yet, since this is a small number of maneuvers which are not precisely in orthogonal directions, the surface deformations are never purely axisymmetric.

\begin{figure*}[t]
\includegraphics{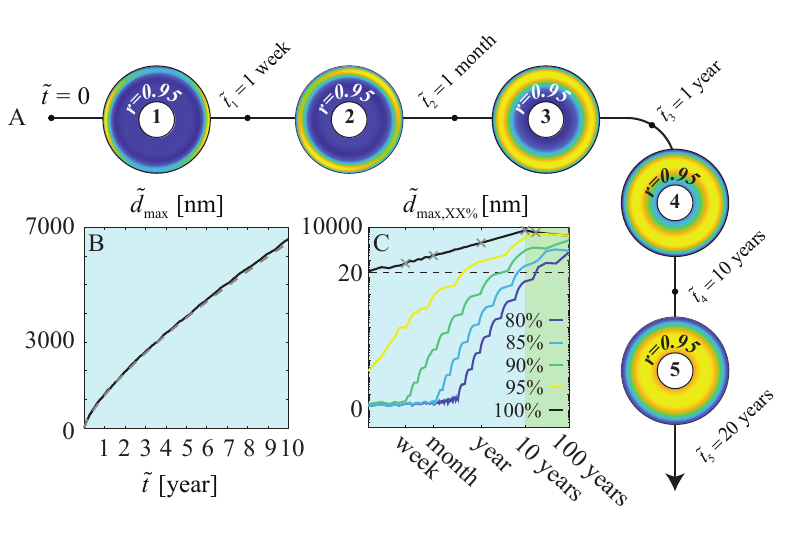}
\caption{Long-term deformation illustration of a fluidic mirror during a 10-year period, featuring a daily 45$^{\circ}$ slewing maneuver around a random in-plane rotation axis. (A) The mirror deformation at various times (A1 – 1 week, A2 – 1 month, A3 – 1 year, A4 – 10 years, A5 – 20 years), showing the deformation becoming increasingly axisymmetric due to the variety of maneuver directions. (B) The maximum deformation across the telescope surface, accumulating to 6400~nm over 10 years (solid black line) and the analytical prediction Eq.~\ref{eq:36} (gray dashed line). (C) The same deformation data on a logarithmic scale, together with the maximal deformations over smaller apertures, and during a hypothetical relaxation period. The deformation exceeds the selected critical threshold of 20~nm at the 95\% radius (yellow line) after approximately six months, while for the 80\% aperture (dark blue line), the maximal deformation remains below 20~nm for more than 20 years. The gray 'x' marks correspond to the time points at which the deformations maps in A are depicted.}
\label{fig:Fig8}
\end{figure*}

To gain a better understanding of mirror surface deformations during operation of a notional fluidic telescope, we studied a basic maneuver construct (Figure~\ref{fig:Fig6}) as well as a typical day's worth of maneuvers (Figure~\ref{fig:Fig7}). However, a realistic telescope would maneuver constantly throughout its lifetime — ideally tens of years. The results thus far clearly indicate that all maneuvers contribute to the deformation of the surface, and that relaxation driven by the surface tension force alone is too slow to recover the original spherical shape within a reasonable time. At the same time, we also see that the deformations propagate very slowly toward the center of the telescope. We next seek to examine the nature of deformations accumulated over a long period of continuous maneuvers and draw guidelines based on the understanding we gain from non-dimensional analysis.

Figure~\ref{fig:Fig8} presents the deformation of a fluidic mirror over an actuation period of 10 years. While a realistic case consists of many small maneuvers each day, previous analysis (Figure~\ref{fig:Fig7}) shows that a simulation of a single large maneuver per day can serve as the worst-case scenario. Thus, to reduce the computational load, in our long-duration simulations we substitute a set of short maneuvers with a single large maneuver of 45$^{\circ}$ per day, around a different random in-plane rotation axis. Figure~\ref{fig:Fig8}A presents the shape of the mirror surface as it evolves over time. Since the interface deformation is constructed of many rotations in different directions, in this case (consisting of 3650 maneuvers of 45$^{\circ}$ each) the deformation approaches an axisymmetric shape. Figure~\ref{fig:Fig8}B shows the maximal deformation over the entire telescope surface, which builds up to 6400~nm during the 10-year operation period. Figure~\ref{fig:Fig8}C presents the same data on a logarithmic scale and includes the maximal deformation for smaller apertures, as well as the behavior in a hypothetical relaxation period. Most importantly (as can also be observed in the colormaps of Figure~\ref{fig:Fig8}A), the deformation passes the threshold of 20~nm at the 95\% and the 80\% radius curves only after 6 months and 20 years respectively — this deformation is the result of the earliest maneuvers that have just now reached this radius.

Interestingly, although the deformation pattern varies in both time and space, the maximal deformation follows the same trend discovered in the non-dimensional analysis for a single maneuver $\tilde{d}_{\text{max}} = AC_{\text{seq}}t^{3/4}\tilde{h}_0$. While in this long-term operation the deformation does not grow monotonically, as predicted theoretically for a single maneuver, the theoretical trend still provides an excellent approximation, as seen in Figure~\ref{fig:Fig8}B and Figure~\ref{fig:Fig8}C (gray dashed line). To quantify this behavior, we introduce the constant $C_{\text{seq}}$, which represents a scaling factor linking the deformation from a single maneuver (obtained from the non-dimensional analysis) to the cumulative deformation from many sequential maneuvers. This constant is defined by considering the fraction of time the telescope spends maneuvering, relative to the total period (actuation plus rest), and by averaging the effect of rotations around various axes.

For a large number of maneuvers, we assume that the resulting deformation is axisymmetric, similar to the outcome of rotating alternately about two perpendicular axes. Thus, for sequential operation, we propose the following expression, $C_{\text{seq}} = C_a \tilde{t}_a / 2\tilde{t}_p$, where $C_a$ is the constant found in the non-dimensional analysis for a single actuation, $\tilde{t}_a$ is the actuation time, and $\tilde{t}_p$ is the full cycle period (actuation + rest). The factor of 2 accounts for the effective averaging over two perpendicular axes of rotation, assuming sufficient statistical coverage of orientation over time. Substituting into our model yields the theoretical prediction for the maximal deformation as a function of the fluidic mirror properties, as well as the slewing dynamics:

\begin{equation}\label{eq:36}
\tilde{d}_{\text{max,seq}} = C_a \frac{\tilde{t}_a}{2\tilde{t}_p} \frac{\tilde{\rho}\tilde{D}\tilde{\omega}^{5/4}\tilde{h}_0^{9/4}}{2\tilde{\gamma}^{1/4}} \left(\frac{\beta}{3\tilde{\mu}}\right)^{3/4}
\end{equation}

This expression captures the fact that deformation accumulates only during active slewing and benefits from the axisymmetric deformation profile resulting from randomly oriented maneuvers over a long period of time.

\section{Concluding remarks}

In this paper we introduced a theoretical model and an analytical solution that describe liquid mirror dynamics for a fluidic space telescope under one of the most significant sources of perturbations — its slewing maneuvers required to change the observation point in space. Based on the model, we identified different regimes for both actuation and relaxation stages, providing valuable insights into the resulting deformation. For example, we observed that the maximal deformation at the actuation and relaxation stages scales as $t^{3/4}$ and $t^{-1/4}$, respectively. Furthermore, we derived simple expressions for the maximal deformation and piston deformation, highlighting their dependence on the liquid mirror properties and the maneuver characteristics. Equation~\ref{eq:33} reveals that the maximal deformation of the film at actuation scales with the film thickness as $\tilde{h}_0^{9/4}$ and with the angular velocity as $\tilde{\omega}_0^{5/4}$. These results can serve as design guidelines for the telescope and its slewing plan.

Additionally, from an optical perspective, not only the maximal deformations matter, but also their spatial distribution. Our simulations predict that for a hypothetical telescope with a 50-meter diameter, 1-mm thick liquid mirror, the maximal deformation could reach several micrometers after 10 years of operation. However, within the inner 40 meters (i.e., 80\% of the aperture), the deformation remains under 20~nm even after 20 years. This difference highlights the importance of a dynamic model, as the viscoelastic timescales can be very long and may strongly affect the functionality of the overall system. These considerations may need to be incorporated in the design process of the telescope, where an obvious degree of freedom is to control the aperture used through an adjustable aperture stop (e.g., a mechanical iris), as is commonly done in many optical systems.

Another potential way of minimizing long-lasting mirror surface disturbances of a fluidic space telescope is through its concept of operations. In creating a sequence of astronomical observations, the expected surface deformation could be combined with other factors (e.g., timing, scientific value, and control effort required) in a multi-objective optimization in order to take maximum advantage of the remaining useful life of the telescope.

Ultimately, after sufficient use, the optical quality of the mirror surface will degrade. A “reset” procedure that restores the entire surface of a fluidic mirror to its original, precise geometry would be highly desirable. One possible option for such a procedure could be to combine linear acceleration and angular velocity around the z-axis, as is done with ground-based liquid-mirror telescopes, shaping the liquid surface into a paraboloid. Other physical mechanisms include electromagnetic actuation similar to those employed in the Zenith program, or leveraging the thermocapillary effect to reshape the liquid and restore its optical properties~\cite{Eshel2022}. 

The aforementioned actuation mechanisms can be incorporated into our finite-domain, thin-film model of fluidic mirrors as different source terms for equation~\ref{eq:9}. However, an additional step is needed when actuating via electromagnetic fields or temperature distributions across the mirror surface. In these cases the mirror surface evolution equation would need to be coupled Maxwell's or an energy balance equations, respectively. In recent work~\cite{Ref27}, we used surface electrodes at the chamber bottom to apply DEP (electrical) interfacial forces to shape the liquid, while introducing the 1D non-self-adjoint thin-film model. We solved the electric field distribution numerically and, assuming small deformations, decoupled the two governing physical phenomena (electrostatics and fluid mechanics). We then used the numerical results as the source term for the thin-film equation. Importantly, we note that DEP stress and magnetic stress on a ferromagnetic fluid are very similar in nature. Both stresses originate from the differences in electric permittivity and magnetic susceptibility between the liquid and the air above it, respectively, and both apply force distributions at the interface in the normal direction. Thus, with the theoretical model suggested for magnetic stress at the interface~\cite{Ref21}, one should be able to readily incorporate this mechanism into our thin-film model. 

\begin{acknowledgments}
This project was funded by the European Union (ERC, Fluidic Shaping, 101044516). Views and opinions expressed are, however, those of the authors only and do not necessarily reflect those of the European Union or the European Research Council Executive Agency. Neither the European Union nor the granting authority can be held responsible for them. We also gratefully acknowledge funding from the Israel Science Foundation grant no. 2263/20. I.G. acknowledges the support of ISEF and the Azrieli Foundation.
\end{acknowledgments}

\bibliographystyle{apsrev4-2}
\bibliography{references} 

\end{document}